\documentclass{article}[12pt]
\title{Dealing with multiple testing: To adjust or not to adjust}
\author{Yudi Pawitan\footnote{Corresponding email: yudi.pawitan@ki.se}\ \  and Arvid Sjölander\\
Department of Medical Epidemiology and Biostatistics\\
Karolinska Institutet}
\date{September 2020}

\newcommand{\Prob}{\mbox{Prob}}

\begin{document}
\maketitle

\begin{quote}
\textbf{Abstract:} Multiple testing problems arise naturally in  scientific studies because of the need to capture or convey more information with more variables. The literature is enormous, but the emphasis is primarily methodological, providing numerous methods with their mathematical justification and practical implementation. Our aim is to highlight the logical issues involved in the application of multiple testing adjustment.
\end{quote}

\section{Introduction}
Let’s start by saying that dealing with multiple comparisons is tricky, perhaps not mathematically, but logically for sure. The wrapping of this problem should be labeled with at least two warnings, one from Oscar Wilde: `the truth is rarely simple and never pure’ and the other from Einstein: `as far as the laws of mathematics refer to reality, they are not certain; and as far as they are certain, they do not refer to reality’. Einstein’s statement was made in a lecture on geometry in relation to experience, partly highlighting his insight that Euclidean geometry fails to explain reality, so any branch of mathematics less intuitively obvious in its relation to reality than geometry has an even larger gap to close.  

Multiple comparisons arise naturally in most scientific studies because of the need to capture or convey more information with more variables. But there are many steps taken during a scientific process that also have multiplicity implications, such as deciding the appropriate scale (linear or log), variables to be included in the analysis, subgroup analysis, etc. 

First, the easy part: mathematically precise statements can be derived \emph{assuming the null hypotheses are true}; in effect we are dealing with random numbers with no signal. At 5\% significance level, for every 100 tests we shall get 5 spuriously significant tests on average. Alternatively, assuming independence and still under the null, it is virtually certain -- probability $=1-(1-0.05)^{100}=0.994$ -- that the most significant result among the 100 tests will have a raw P-value  $\le 0.05$. This means that if we are naïve we would be easily misled by false positives, hence the usual warning about data dredging.  

In practice, to account for multiplicity, one can simply adjust the P-values by multiplying each by the number of tests (say $M$), or dividing the nominal significance level by $M$. This makes it harder to declare significance. Using this so-called Bonferroni correction, we can guarantee that if we only reject hypotheses with adjusted P-value $\le 0.05$, say, then the probability of making any false rejection among all tests is $\le 0.05.$ To prove this, suppose we have $M$ null hypotheses, where a subset $M_0$ are true; let $F$ be the number of false rejections. Then  
\begin{eqnarray*}
\Prob(F>0) &=& \Prob(\mbox{adjusted P} \le 0.05,\ \mbox{at least one of } M_0 \mbox{ hypotheses})\\ 
&=&\Prob(\mbox{P-value} \le 0.05/M, \mbox{ at least one of } M_0 \mbox{ hypotheses})\\ 
&\le& M_0\times 0.05/M\\  
 &\le&  0.05.      
\end{eqnarray*}

Due to its simplicity, this procedure is commonly used in practice. If we want to protect ourselves against false positives, it seems reasonable to apply this procedure. The literature on multiple testing is enormous and there are numerous alternative methods, many as improvements of the Bonferroni correction, indicating that multiple-testing correction is taken seriously, and when applied, it is mathematically straightforward. 

\begin{quote}
\textbf{Example 1: Cancer drug study.} The following table shows the P-values in a study to compare a metastatic cancer drug vs placebo for 10 patient characteristics. The best raw P-value is 0.007 for the Karnofsky index (let’s keep it mysterious for now), but when corrected using the Bonferroni method, nothing is significant at 5\% level. The last column actually gives an estimated number of false positives when we consider the variable and those above it as significant, so we do not truncate it at 1. We will use this quantity later.
\end{quote}

\begin{table}[h!]
    \centering
    \begin{tabular}{llcc}
& Variables    & P-value  & 10$\times$P-value \\
\hline
1 & Karnofsky index  &   0.007   &    0.07 \\
2& Body weight   &    0.013  &   0.13 \\
3 &Tricep skin-fold   &    0.091   &   0.91 \\
4& Hemoglobin concentration  & 0.236  &  2.36 \\
5& Erythr.\ sedimentation rate  &   0.350  &    3.50 \\
6 &Albumin in serum  &  0.525  &   5.25 \\
7& Creatinine in serum   &     0.535   &   5.35 \\
8 &Bilirubin in serum  &  0.662    &  6.62 \\
9& S-alkalinephosphatase   &   0.823   &   8.23 \\
10& Alanine aminotransferase  &  0.908   &   9.08 \\
\hline
    \end{tabular}
    \caption{\em P-values in a study to compare a metastatic cancer drug vs placebo for 10 patient characteristics. The third column is the Bonferroni-adjusted P-value that is (on purpose) not truncated at 1.}
    \label{tab:drug}
\end{table}

\section{Not everyone wants to adjust}

Yet the statistical concerns about false positives are not shared universally among scientists. The clearest objection was formulated by Rothman (1990), who declared ‘No Adjustments are Needed for Multiple Comparisons’. Being the first Editor of the journal \emph{Epidemiology} and author of a widely used textbook, Rothman is one of the most influential and highly cited epidemiologists, so it is worth understanding his arguments. Briefly, two non-mathematical presumptions are needed for application of multiplicity adjustments: 
\begin{itemize}
\item the `universal null hypothesis’, covering all hypotheses under consideration, is a reasonable state of nature, so that chance does cause many unexpected findings, and 
\item no one would want to investigate further something that could be caused by chance.
\end{itemize}
He argued that in a scientific process these presumptions are not true, that `chance’ is not a scientific explanation, so scientists should `grasp at every opportunity’ to understand unusual findings, and that the possibility of being misled is part of the trial-and-error process of science.  

It is still the case today that most statistical results in epidemiology and medical literature are rarely adjusted for multiple comparisons, with notable exceptions in clinical trials and high-throughput molecular studies; see more below. In clinical trials, multiplicity issues arise in, for example, the choice of hypotheses to be tested, sidedness of the test, interim analyses during the trial, main analysis plan, subgroup analyses after the trial, etc. 

The US Food and Drug Administration’s \emph{Guidance for Industry: E9 Statistical Principles for Clinical Trials} stipulates that they should be addressed explicitly in advance. In particular, for multiple comparisons, `adjustment should always be considered and the details of any adjustment procedure or an explanation of why adjustment is not thought to be necessary should be set out in the analysis plan.' 

Furthermore, since September 2007, in the so-called FDAAA 801 Requirement, any clinical trial of drugs or medical interventions that will seek FDA approval must be registered when the trial begins (see: https://www.clinicaltrials.gov), hence limiting or avoiding completely the reporting of unplanned analyses. Why are the strict guidelines not universally adopted in science in general? Imagine asking scientists to register all the hypotheses and analysis plans -– including interim analyses -- in advance. 

\begin{quote}
\textbf{Example 1: Cancer drug study (continued).} In the cancer drug study, suppose we know nothing about the variables prior to the study, so for us all these variables assume equal status. Then it would be naive to take the raw P-values seriously, thus forcing us to accept the adjustment and the overall null result. But suppose prior to collecting the data, because this is a study on end-stage metastatic patients, we declared that the Karnofsky index, a generalized measure of functional performance, was of primary interest, while the other variables were of secondary interest. Then, no adjustment would be necessary for the Karnofsky index. Thus, as accepted in clinical trials, in assessment of evidence \emph{our intention matters}. This does not feel controversial. We note that, for justifying the non adjustment, the primary interest in the index does not require prior knowledge or data or anything sensible; in principle we only need to declare it in advance.  
\end{quote}

Now suppose we accept we know nothing prior to the study, hence the multiplicity adjustment, but another research group who is working on the Karnofsky index contacts us to share the data. We agree to give them that variable only. Now, for them, it seems reasonable that no adjustment is needed and to conclude that the Karnofsky index is statistically significant. So, here we have a seeming paradox that, with the same data, two research groups can claim different evidence: one group cannot claim significance, but the other can. But what if we contact them? Now it seems the adjustment must be used, since we could have contacted any group working with the most significant variable. This means the mechanism of contact becomes an issue, but in practice the appearance of the ‘second group’ in the scene can of course be completely haphazard, e.g. via a chance encounter at a party, a friend of a friend, etc. In this social contact, how do we keep track of who brings up the topic first? It would be impossible to formalize such a process. 

\section{Single test} 
Surprisingly, the logical issue associated with the application of multiplicity adjustment arises even when we only perform a single test (Berger and Berry, 1987). Suppose a client comes to a statistician with a study (say with sample $n=100$), and the statistician performs a single test and obtains $z= 2.1$ (P-value = 0.036). It seems uncontroversial to claim significance at $\alpha=0.05.$ Yes, but wait … what did he plan to do if the result was not significant? Suppose, as with other scientists, he planned to collect more data. So, his actual procedure is a sequential test as follows: 
\begin{enumerate}
\item Collect $n=100$ samples and test if $|z_1| > c$, if significant stop.  
\item   Otherwise, collect 100 more samples, and test if the overall $|z_2| > c$. 
\end{enumerate}
To get an overall significance level $\alpha=0.05.$, we must use $c=2.18$, such that 
$$
\Prob(|z_1| > c) + \Prob(|z_1| \le c \mbox{ and } |z_2| > c) = 0.05. 
$$
So, the observed $z=2.1$ is actually not significant! Again, the statistical significance is affected not just by the data, but also by the intention of the scientist, but in this case it feels disturbing because the second stage of the study is still only a thought. (Before the reader accuses this example as perverse, adjusting for the intention here follows the FDA guideline on interim analysis of clinical trials.) 

\section{Logical difficulties, Type-I and II errors}

These examples highlight a generic logical question: what collection of tests do we want to apply the adjustment to? To cover all tests relating to the same biological process? All tests in a single paper? If the latter, then theoretically we can avoid multiplicity correction by splitting the results into separate papers. The primary problem is that we are using the same statistic (P-value) both as a measure of evidence in a specific dataset (statistical distance between the hypothesis and the data) and as a measure of uncertainty (decision-making error rates) over hypothetical repetitions of the study. 

In the former, adjustment for multiplicity is not an issue, but in the latter it is. However, the latter requires a precise setup of how the study repetitions are to be done, and here \emph{intention matters}, for example, in deciding which tests are to be included or prioritized. The examples show that any test can legitimately belong to distinct collections with distinct repetition studies that depend on the perspective of the experimenters. (This is not strange, e.g. a person may belong to distinct clubs with conflicting rules.) In the drug example, different groups of researchers are imagining different hypothetical experiments involving the Karnofsky index. In the single-test example, on seeing the observed data, the statistician’s first reaction is to imagine hypothetical repetitions involve one test, but the scientist’s intention involves hypothetical repetitions with two tests. Whose perspective is correct?  

In view of this logical difficulty, how are we to react to potentially conflicting conclusions from unadjusted and adjusted analyses? Actually, we are unwittingly put into this corner by an implicit demand that we make a decision. When a study is only performed once, i.e. the hypothetical repetitions remain hypothetical, we are in a never-ending unsolvable logical puzzle on how to decide. How do we break this puzzle? 

In clinical trials, for the key hypothesis, we break it simply by decree (e.g. following the FDA Guidelines), in essence limiting or avoiding the issue by stating the hypothesis and analysis methods in advance. Science does not follow such strict rules, but we need validation studies to confirm interesting discoveries. Before further confirmation discoveries are considered provisional, so, in contrast to clinical trials, \emph{it is not necessary to make a decision} about the true state of nature. However, we also note that, to confirm a discovery, it is not necessary to perform exactly the same experiment as before (i.e. the hypothetical repetitions). For example, a discovery in an observational study in human will be substantially more credible if validated biologically in mice, and vice versa.  

Eventually, in treating a study as a screening tool to identify interesting discoveries, we have to go back to the basic trade-off between type-I (false positive) and type-II (false negative) errors: protecting against one will increase the other. Strict adherence to multiple-testing adjustment protects against inflation of type-I errors and increases type-II errors, but who decides which error is more important? The legendary investor George Soros reportedly said, ``It's not whether you're right or wrong [that is important], but how much money you make when you're right and how much you lose when you're wrong.'' It does not seem to make sense to be feel strongly against one type of error vs the other independently of the context.

Different areas of science may treat multiplicity adjustment differently, perhaps depending on the relative costs of the errors, their history/experience with the errors and abundance of potential leads for discoveries. For example, molecular epidemiology went through the lamented candidate-gene approach to complex diseases, roughly from 1980s up to early 2000s (Hirshhorn et al, 2002; Chabris et al, 2012), where few findings were replicated. 

\begin{quote}
\textbf{Example 2: Publication bias and Winner’s curse.} The human genome is a rich source of variables/genes. For many complex phenotypes, suppose there is no real effect or, more likely, the individual-gene effects are so tiny that the power of fundable studies is too small to detect anything. Say there are 100 research groups investigating different genes and phenotypes; each group is essentially generating random numbers. At 5\% level, there will be 5 lucky groups with significant results: these are much more likely to get published, and then fail to replicate. If these are really 100 distinct groups, what can we do about this problem? No system of publication now can communicate so many negative findings to balance the false positives, so the problem seems to be an inevitable price of the scientific process.
\end{quote}

When high-throughput molecular studies, particularly the genome-wide association studies (GWAS) came into the scene, a single study/paper routinely performs millions of tests of single nucleotide polymorphisms (SNPs), and the genome-wide significance threshold $5\times 10^{-8}$ based on the Bonferroni correction became the accepted method of dealing with the huge multiplicity problem. One may argue correctly that we would be missing a lot of signals (type-II errors), but since the field had seen a lot of wasted effort at replicating false leads during the candidate-gene era, the consensus on the use of Bonferroni correction seems unchallenged. 

The problem in molecular studies in the genomic era is that there are simply too many potential leads, so one needs a method to limit them. Note, however, that the P-values for each phenotype are usually adjusted separately, e.g. if there are 1M SNPs, then the $5\times 10^{-8}$ is applied for each phenotype regardless of the number of phenotypes, so the multiplicity adjustment is not followed consistently. 

\section{Flexible multiplicity adjustment}

Multiple testing ideas are useful during an exploratory phase of a study. Goeman and Solari (2011) identified three sensible requirements for a multiple correction procedure: 
\begin{itemize}
\item not too strict: should allow possibilities of false rejections 
\item post-hoc: should allow choice after seeing the data
\item flexible: should allow whatever results to pursue, not just the significant ones. 
\end{itemize}

Instead of focusing on the probability of making any false rejections, which is too strict, we can instead estimate and provide confidence lower-bounds for the number of true discoveries (= correct rejections). To emphasize its post-hoc feature, they called their procedure `cherry picking’. Recent developments, for example in false discovery rate (FDR) estimation, are also in line with these requirements (e.g. Lee et al, 2012). The purpose is more to set realistic expectations for further studies rather than making final decisions about the true state of nature.  

\begin{quote}
\textbf{Example 1: Cancer drug study (continued).} Suppose this study was performed on end-stage cachexic patients, which are characterized by severe wasting/loss of body mass and functional impairment. The top 3 variables are then of special interest, but, crucially, this interest can be decided post-hoc \emph{(after seeing the data)}. From the third column (10$\times$P-value), the estimated number of true discoveries is $3-0.91 \approx2$ (Lee et al, 2012). An application of the cherry-picking procedure (Goeman and Solari, 2011) gives 95\% guarantee of at least 1 true difference from the top 3 findings, and 75\% guarantee of at least 1 true difference from the second and third tests. So, even if we do not want to specify any hypothesis in advance, the analysis indicates the drug is worth studying further.
\end{quote}

\section{Conclusions}
To conclude, we do not at all mean to sound skeptical of the use of multiplicity adjustment, especially when put in a flexible false discovery rate framework. However, we do want to emphasize the nuances when adjustment is applied to real studies, and the constant need to consider the type-I and type-II errors, whose relative costs are highly context specific. We highlight some logical problems with its formal use when assessing scientific evidence, thus partly explain the lack of universal acceptance. In areas of science where potential leads for discoveries are not abundant and the true discoveries are rare but fundamental, formal multiplicity adjustment is not the norm. On the other hand, there are areas that must pay close attention to the multiplicity adjustment, e.g. clinical trials, where type-I errors can have enormous human or financial cost, or high-throughput molecular studies with the flood of false positives when we use traditional significance level. 

Nature does not reveal its truths easily. Anyone in search of those ultimate truths will tolerate certain false positives along the way. Healthy skepticism is a necessary default attitude in the error-prone but self-correcting process of science. But there is a legitimate concern of false positives when a scientific result is disseminated to the public. The scientists and the public may have different relative costs of the type-I and type-II errors: the scientists may be more concerned with missing a discovery (type-II), but the public does not like to be misled by false positives. Also, the public may not have the patience or the kind of skepticism that accepts results as provisional. They are like observers that do not have sufficient background knowledge -- and attention span -- to absorb conflicting accounts of an event faraway. So, conflicting findings, say, about the effects of electromagnetic field or coffee or cancer screening, etc., will confuse the public. The public position is similar to the one we mention above, where we put ourselves unnecessarily into a corner by thinking that we have to decide one way or the other. The public wants to know `the truth' one way or the other, but the scientific answer comes in confusing bits. The scientists themselves can live with the uncertainty of a provisional result or with a variegated truth about effects. It is not obvious how to reconcile the scientific acceptance of complex uncertainties and the public need for a simple message. 

\section*{Acknowledgement}
An earlier version of this manuscript was published in \emph{Qvintensen} (Nr 2, 2014), a bulletin of the Swedish  Statistical Society. We are grateful to the Editor Dan Hedlin who arranged the publication and kindly helped in editing the manuscript. This is an accompanying manuscript to \emph{Defending the P-value}, which is also published in the Arxiv.

\section*{References}
\begin{description}
\item Berger, JO and Berry, D (1987). The relevance of stopping rules in statistical inference. In \emph{Statistical  Decision Theory and Related Topics} IV (S Gupta and J Berger, Eds). New York: Springer Verlag. 

\item Chabris CF, et al. (2012). Most reported genetic associations with general intelligence are probably false positives. \emph{Psychol Sci.} 23(11): 1314-23. 

\item Goeman J and Solari A (2011). Multiple testing for exploratory research, with discussion. \emph{Statistical Science}, 26 (4), 584-597. 

\item Hirschhorn JN, et al. (2002). A comprehensive review of genetic association studies. \emph{Genetics in Medicine} 4(2): 45-61. 

\item Lee W, et al. (2012). Estimating the number of true discoveries in genome-wide association studies. \emph{Statistics in Medicine}, 31(11-12): 1177-89 

\item Pawitan Y (2001). \emph{In All Likelihood}. Oxford: Oxford University Press. 

\item Rothman, KJ (1990). No Adjustments are Needed for Multiple Comparisons. \emph{Epidemiology}, 1 (1), 43-46.  
\end{description}

\end{document}